\documentstyle[prc,preprint,tighten,aps]{revtex}
\begin {document}
\draft
\title{New representation of orbital motion with arbitrary angular momenta}
\author{Y. Suzuki$^{1}$, J. Usukura$^{2}$, and K. Varga$^{1,3}$}
\address{$^1$Department of Physics, Faculty of Science, Niigata University,
Niigata 950--21, Japan\\
$^2$Graduate School of Science and Technology, Niigata University,
Niigata 950--21, Japan\\
$^3$Institute of Nuclear Research of the Hungarian
Academy of Sciences, Debrecen, H--4001, Hungary}
\date{\today}

\maketitle

\begin{abstract}
A new formulation is presented for a variational calculation of 
$N$-body systems on a correlated Gaussian basis 
with arbitrary angular momenta. The rotational motion of the system 
is described with a single spherical harmonic of the total angular 
momentum $L$, and thereby needs no explicit coupling of 
partial waves between particles. A simple generating function for the 
correlated Gaussian is exploited to derive the matrix elements. 
The formulation is applied to various  
Coulomb three-body systems such as $e^-e^-e^+,\, tt\mu,\, td\mu$, 
and $\alpha e^-e^-$ up to $L=4$ in order to show its usefulness 
and versatility. A stochastic selection of the basis functions yields 
good results for various angular momentum states. 
\end{abstract}

\narrowtext
\section{Introduction}
We have recently shown \cite{prca} that the stochastic 
selection of the correlated Gaussian \cite{corrgauss,temper} 
basis functions leads to precise variational solutions for diverse 
fermionic and bosonic nonadiabatic $N=2-7$-body systems. 
The trial function in a variational approach must be flexible enough 
to describe 
the full variety of correlations between the particles. The correlation is 
conveniently represented by a correlation factor, $F=\prod_{i<j}^{N}  
f_{ij}$, and thus the trial function is often 
chosen to be of this form. The trial function of Hylleraas type,  
which is used in atomic and molecular physics, approximates $f_{ij}$ 
as a linear combination of exponentials, ${\rm exp}[-\alpha_{ij} 
\vert {\bf r}_i - {\bf r}_j \vert ]$.  If $f_{ij}$ is 
approximated as a linear combination of Gaussians, ${\rm exp}[-\alpha_{ij} 
({\bf r}_i - {\bf r}_j)^2]$, the $N$-particle trial 
function then contains a product of these Gaussians: 
$\prod_{i<j}^{N}{\rm exp}[-\alpha_{ij}({\bf r}_i-{\bf r}_j)^2]=
{\rm exp}[-\sum_{i<j}^{N}\alpha_{ij}({\bf r}_i-{\bf r}_j)^2]$ 
\cite{KA91,KP93,cencek}. 

\par\indent
The above form of the correlated function describes the motion with the 
orbital angular momentum $L=0$ only. It is obviously important to extend the correlated function 
to the case with $L>0$, and in fact  there is an 
increasing interest in finding a precise solution with $L\ge 2$ for 
the Coulomb three-body systems \cite{frolov96}.   
The standard way to describe the rotational motion is to vectorially couple 
the solid spherical harmonics of the relative coordinates 
\cite{frolov96,kamimura}. Each of the 
solid spherical harmonics carries the partial wave for the corresponding 
relative motion. Since these partial waves are in general not good quantum 
numbers, 
several sets of partial waves are in general necessary for a realistic 
description of the motion, particularly in a nuclear system 
\cite{kameyama,vos}. Another way to incorporate the angular dependence is 
to use Cartesian Gaussian functions. The calculation of matrix elements  
becomes complicated in both cases, especially when the number of 
particles increases or high angular momenta are involved.  
A quite different way to introduce the angular part, proposed in 
\cite{prca}, involves no 
partial wave decomposition for each relative motion but attempts to 
determine a vector which has closest relevance to the rotational 
motion. The vector is defined as a linear combination of the relative 
coordinates and their coefficients can be treated as variational 
parameters. The formulation using this new angular part is entirely 
free from the complexity involved in the angular momentum coupling.        

\par\indent
The purpose of this paper is to exploit this formulation in greater detail, 
to present the formulas needed for 
$N$-body system interacting via central force, and to test its utility by 
applying to Coulomb three-body problems such as the muonic molecule 
and the helium atom.  Several authors have investigated both the ground 
and excited states of these systems, for example in 
\cite{kamimura,vinit,bhatia,alexander,frolov94} for the muonic 
molecule and in \cite{frolov86,sims} for the helium atom.  A comparison 
with the solutions known in literature 
will be useful to judge the utility of the present formulation. Although 
any spherically symmetric orbital functions can be used together with 
the angular part introduced here, the correlated Gaussian has the advantage 
of simplicity in the coordinate transformation.   

In section II we introduce the correlated Gaussian with the angular 
momentum dependence and show that the matrix elements can easily be 
evaluated with the use of the generating function of the correlated 
Gaussian. No problem arises as to the center-of-mass motion 
because the formalism does not include any dependence on the center-of-mass 
variable. In section III some of the results for the Coulomb three-body 
system are presented within the new formulation for the angular 
dependence and compared to those available in literature. A summary 
is given in section IV.  

\section{Formalism}
\subsection{The correlated Gaussians}

Any square-integrable function with angular momentum 
$lm$ can be approximated, to any desired accuracy, by a linear combination 
of nodeless harmonic-oscillator functions (Gaussians) of continuous size 
parameter $a$ \cite{bukowski}:
\begin{equation}
\Gamma_{lm}({\bf r}) \sim {\rm e}^{-{\frac{1}{2}} a r^2}
{\cal Y}_{lm}({\bf r}),\ \ \ {\rm with}\ \ \ {\cal Y}_{lm}({\bf r})=
\vert {\bf r} \vert^l Y_{lm}(\hat{\bf r}).
\end{equation}
A generalization of this to $N$-particle systems contains a product 
of the Gaussians 
${\rm exp}[-\sum_{i<j}^{N}\alpha_{ij}({\bf r}_i-{\bf r}_j)^2]$. 
The product can be conveniently expressed in 
terms of a set of $(N-1)$ independent relative  
coordinates ${\bf x}$, $({\bf x}_1,...,{\bf x}_{N-1})$, 
instead of $N(N-1)/2$ interparticle distance vectors 
$({\bf r}_i-{\bf r}_j)$. By a set of relative coordinates we mean 
the one in which the intrinsic kinetic energy operator takes the form 
$\sum_{i=1}^{N-1}{\bf p}_i^{2}/2\mu_i$ with reduced masses $\mu_i$. 
Even with this condition there are a number of possible sets of 
relative coordinates. One can choose, however, any one of the sets as 
${\bf x}$ because each set of relative coordinates is obtained from 
any other set of relative coordinates by an appropriate $(N-1)\times(N-1)$  
matrix $T$. The relative coordinates are assumed to be normalized 
in such a way that the volume element remains 
unchanged under the coordinate transformation, which requires that 
the determinant of $T$ is unity. An $N$-particle basis function, the 
so-called correlated Gaussian, then looks like
\begin{equation}
\psi_{LM}(A, {\bf x})=
{\rm e}^{-{1 \over 2}\tilde{{\bf x}}A {\bf x}} \theta_{LM}({\bf x}),
\end{equation}
where $A$ is an $(N-1)\times(N-1)$ positive-definite, 
symmetric matrix containing $N(N-1)/2$ nonlinear parameters, specific 
to each basis element, 
and the quadratic form, $\tilde{{\bf x}}A{\bf x}$, involves scalar products 
of the Cartesian vectors: $\tilde{{\bf x}} A {\bf x}=\sum_{i=1}^{N-1} 
\sum_{j=1}^{N-1}A_{ij}{\bf x}_i\cdot{\bf x}_j$.

\par\indent
The function $\theta_{LM}({\bf x})$ in Eq. (2), which represents the 
angular part of the wave function with the total orbital angular 
momentum $L$ and its projection $M$, is a generalization of 
$\cal Y$ in Eq. (1).  Usually it is chosen as a vector-coupled 
product of solid spherical harmonics of the relative coordinates
\begin{equation}
\theta_{LM}({\bf x})=\left[[[{\cal Y}_{l_1}({\bf x}_1){\cal Y}_{l_2}
({\bf x}_2)]_{L_{12}} 
{\cal Y}_{l_3}({\bf x}_3)]_{L_{123}}...\right]_{LM} ,
\end{equation}
where the square bracket stands for the coupling of angular momenta. 
Each relative motion has a definite angular momentum in Eq. (3). Since the 
set of angular momenta itself is not a conserved quantity, it may be 
important to include several sets of angular 
momenta $(l_1,l_2,...,l_{N-1};L_{12},L_{123},...)$ for a realistic 
description.  This is the case especially in nuclear few-body 
problems \cite{kameyama,vos,SVAO}. It is also noted that a faster 
convergence is in general obtained by allowing the use of different 
sets of relative coordinates together with suitable sets of angular momenta. 
From the fact that $\theta_{LM}({\bf x})$ can be expressed by different 
partial wave decompositions in different relative coordinate systems, 
one can conclude that the usage of partial waves may not be so important 
after all. Besides, 
the various possible partial wave contributions increase the basis 
dimension.  Moreover, the calculation of matrix elements for this choice 
of $\theta_{LM}({\bf x})$ sooner or later becomes too complicated. This 
choice is 
therefore apparently inconvenient especially as the number of particles 
increases and/or the different sets of relative coordinates are employed. 
 
\par\indent
As proposed in \cite{prca}, this difficulty can be avoided by adopting 
a different choice for $\theta_{LM}({\bf x})$:
\begin{equation}
\theta_{LM}({\bf x})=\eta_{KLM}(u,{\bf x})=\vert {\bf v} \vert^{2K+L} Y_{LM}
({\hat{\bf v}}),\ \ \ \ {\rm with}\ \ \  
{\bf v}=\sum_{i=1}^{N-1} u_i {\bf x}_i=\tilde{u}{\bf x}.
\end{equation}
Only the total orbital angular momentum, which is (at least approximately) 
a good quantum number in most cases, appears in this expression. 
The real vector $\tilde{u}=(u_1,...,u_{N-1})$ defines a global 
vector, ${\bf v}$, a linear combination of the relative coordinates, 
and the wave function of the system is expanded in terms of its angle 
$\hat{\bf v}$. The vector $u$ may be considered a variational parameter 
and one may try to minimize the energy functional with respect to it. 
The energy minimization then amounts to 
finding the most suitable angle or a linear combination of angles. 
The continuity of the parameter $u$ can be more advantageous in a 
variational calculation than the discrete nature of the set of the 
angular momenta 
$(l_1,l_2,...,l_{N-1};L_{12},L_{123},...)$ because the change of the 
energy functional can be continuously seen in the former case. The 
factor of $\vert{\bf v}\vert^{2K+L}$ plays an important role in improving 
the short-range behavior of the wave function, e.g., the Coulomb 
cusp ratio \cite{chong}. 
A remarkable advantage of this form of  
$\theta_{LM}({\bf x})$ is that the calculation of matrix elements becomes 
much simpler than in the former case because the coupling of $(N\!-\!1)$ 
angular momenta is completely avoided.

\par\indent
The two forms of $\theta_{LM}({\bf x})$ are in fact closely related to  
each other. It is easy to see that any of the functions of Eq. (4) is 
a linear combination of the terms of Eq. (3), each multiplied by an 
appropriate monomial of the variables, ${{\bf x}_1}^2,...,{{\bf x}_{N-1}}
^2$. For example, it takes 
a particularly simple form for the three-body system, i.e., for the vector 
${\bf v}=u_1{\bf x}_1+u_2{\bf x}_2$  
\begin{eqnarray}
& &\vert {\bf v} \vert^{2K+L}Y_{LM}(\hat{\bf v})=
\sqrt{4\pi}\sum_{k_1,l_1,k_2,l_2 \ge 0 \atop 
2k_1+l_1+2k_2+l_2=2K+L} u_1^{2k_1+l_1} u_2^{2k_2+l_2} \nonumber \\
&\times &\frac{(2K)!!(2K+2L+1)!!}{(2k_1)!!(2k_1+2l_1+1)!!
(2k_2)!!(2k_2+2l_2+1)!!}\sqrt{(2l_1+1)(2l_2+1)\over2L+1}\langle l_10l_20
\vert L0 \rangle \nonumber \\
&\times& \vert {\bf x}_1\vert^{2k_1}\vert {\bf x}_2\vert
^{2k_2}\Big[{\cal Y}_{l_1}({\bf x}_1) 
{\cal Y}_{l_2}({\bf x}_2)\Big]_{LM}. 
\end{eqnarray}  
Table I lists possible sets of $k_1, l_1, 
k_2$, and $l_2$ values for small $K$ and $L$ values. In the case of $K=0$ 
both $k_1$ and $k_2$ are limited to zero and only the stretched coupling, 
namely $l_1+l_2=L$, is allowed. With an increasing 
$K$ value the possible values of partial waves $l_1$ and $l_2$ increase 
including the case of non-stretched coupling. 
To increase $K$ is thus one way to include higher partial waves in 
the calculation. Note, however, that even with $K=0$ additional and 
important partial wave contribution 
comes from the cross term of the exponential part of the correlated 
Gaussian if $A$ is not diagonal.
Conversely any two-variable functions of Eq. (3) with natural parity, i.e., 
$(-1)^{l_1+l_2}=(-1)^L$, may be expressed in terms of 
a linear combination of the terms, $\vert {\bf v} \vert^{2K+L} 
Y_{LM}({\hat{\bf v}})$, by using some appropriate sets of $u$ values,  
each multiplied by a monomial of degree $l_1+l_2-2K-L$ in  
${{\bf x}_1}^2$ and ${{\bf x}_2}^2$. See Appendix for the details. 
A generalization of the argument in Appendix will lead to a conclusion that 
some functions of Eq. (3) may be given in terms of a linear combination of 
Eq. (4), each multiplied by the terms such as ${{\bf x}_i}^2$ and 
$({\bf x}_i\cdot{\bf x}_j)$. Namely, the rotational property of the 
function of Eq. (3) may be represented by combinations of simple forms of 
Eq. (4). Therefore, if one 
can calculate the matrix elements using $\theta_{LM}({\bf x})$ defined 
in Eq. (4), then those with the previous form of $\theta_{LM}({\bf x})$ 
of certain class can be obtained readily. 

\par\indent
The correlated Gaussian we proposed is thus given by 
\begin{equation}
f_{KLM}(u,A,{\bf x})=\eta_{KLM} (u,{\bf x}){\rm e}^{-{1\over 2}
\tilde{{\bf x}} A {\bf x}}. 
\end{equation}
A useful property of the function $f$ is its form-invariance 
with respect to the transformation of the coordinates $\bf x$ to any other 
set of coordinates $\bf y$, that is, for ${\bf x} = T {\bf y}$,
\begin{equation}
f_{KLM}(u,A,{\bf x})=f_{KLM}(u',A',{\bf y}),
\end{equation}
with
\begin{equation}
A'=\tilde{T}AT, \ \ \ \ \ u'=\tilde{T}u.
\end{equation}
Here $\tilde{T}$ is the transpose of $T$. As will be seen later, this 
property is fully exploited in evaluating the matrix elements. It is 
necessary to impose a proper permutation symmetry on the basis states 
for the system of identical particles. The symmetry requirement causes 
a linear transformation of the coordinates, and it can easily be 
incorporated in the present formulation thanks to the form-invariance of the 
correlated Gaussian mentioned above. It would be rather complicated to 
construct the symmetry-adapted basis states using the 
angular function of Eq. (3).

The calculation of the matrix elements 
becomes simpler if one uses a generating function of the correlated 
Gaussian. In fact, the following function $g$ is found to 
be convenient to generate the function $f$ \cite{prca}:   
\begin{equation}
f_{KLM}(u,A,{\bf x})={1\over B_{KL}}
\int d{\hat {\bf t}} Y_{LM}({\hat {\bf t}}) 
\left( {d^{2K+L} \over d\alpha^{2K+L}}
g(\alpha,{\bf t}; u,A,{\bf x})
\right)_{\alpha=0 \atop t=|{\bf t}|=1},
\end{equation}
where
\begin{equation}
g(\alpha,{\bf t}; u,A,{\bf x})=
{\rm e}^{-{1\over 2}\tilde{{\bf x}}A{\bf x}+\alpha {\bf t}\cdot 
(\tilde{u}{\bf x})},
\end{equation}
\begin{equation}
B_{nl}={4 \pi (2n+l)! \over 2^n n! (2n+2l+1)!!}.
\end{equation}
Here $\bf t$ is a unit vector. Equation (9) is easily proved by using 
the simple formula
\begin{equation}
({\bf a}\cdot{\bf b})^{k}=\vert {\bf a} \vert^k \, \vert {\bf b} \vert^k 
\sum_{n,l \ge 0 \atop 2n+l=k} B_{nl} \sum_{m=-l}^{l} 
Y_{lm}({\hat {\bf a}})
Y_{lm}({\hat {\bf b}})^{\ast}. 
\end{equation}

\par\indent
The correlated Gaussian basis of Eq. (6) has parity $(-1)^L$. To construct 
a function with parity $(-1)^{L+1}$, the angular part of Eq. (6) must 
be slightly modified, e.g., to   
\begin{equation}
\theta_{LM}({\bf x})=[\eta_{KL}(u,{\bf x})\eta_{01}(u',{\bf x})]_{LM}.
\end{equation}  
In this case the generating function $g$ of Eq. (10) must be modified 
to include another factor $\alpha'{\bf t}'\cdot (\tilde{u'}{\bf x})$ . 
Equation (9) is then extended to 
\begin{eqnarray}
& &[\eta_{KL}(u,{\bf x})\eta_{01}(u',{\bf x})]_{LM}
{\rm e}^{-{1\over 2}\tilde{{\bf x}} A {\bf x}} \nonumber \\
&=&\frac{1}{B_{KL}B_{01}} \int\!\int d{\hat {\bf t}}d{\hat {\bf t}}'
\Big[Y_{L}({\hat {\bf t}}) Y_{1}({\hat {\bf t}}')\Big]_{LM}
\left( {d^{2K+L+1} \over d\alpha^{2K+L}d\alpha'}
{\rm e}^{-{1\over 2}\tilde{{\bf x}} A {\bf x} +\alpha {\bf t}\cdot(\tilde{u}
{\bf x})
+\alpha '{\bf t}' \cdot (\tilde{u'}{\bf x})}\right)_{\alpha=\alpha'=0 
\atop t=t'=1}.
\end{eqnarray}

\subsection{Calculation of the matrix elements}
In this subsection we will give the details of the method of calculating the 
matrix elements between the basis function of Eq. (6). One can already 
see some useful formulas for $L=0$ motion in \cite{KA91,cencek}. The 
aim here is to demonstrate that the matrix elements for arbitrary $L$ can 
be obtained as simply as those for $L=0$.  As will be seen later, one 
generally needs the matrix elements of the two functions 
which are expressed in terms of different sets of relative coordinates. 

\par\indent
The overlap matrix element is obtained as     
\begin{eqnarray}
& &\langle f_{KLM}(u,A,{\bf y})\vert f_{K'LM}(v,B,{\bf x})\rangle = 
  \langle f_{KLM}(u,A,{\bf y})\vert f_{K'LM}(v',B',{\bf y})\rangle
\nonumber \\
& = & \frac{1}{B_{KL}B_{K'L}} \int\!\int d{\hat {\bf t}}d{\hat {\bf t}}'
Y_{LM}({\hat {\bf t}})^{*} Y_{LM}({\hat {\bf t}}') \nonumber \\
%\left( {d^{\kappa+\kappa'} \over d\alpha^{\kappa}d\alpha'^{\kappa'}}
&\ &\times \left( {d^{\kappa+\kappa'} \over d\alpha^{\kappa}
d\alpha'^{\kappa'}} \,\
\left({(2\pi)^{N-1} \over {\rm det}C}\right)^{3/2}
{\rm e}^{p\alpha^2+p'\alpha'^2+q\alpha\alpha' {\bf t}\cdot{\bf t'}}
\right)_{\alpha=\alpha'=0 \atop t=t'=1},
\end{eqnarray}
with an abbreviation 
\begin{equation}
\kappa=2K+L, \ \ \ \ \kappa'=2K'+L,
\end{equation}
where the property of Eq. (7) is used and where 
\begin{equation}
C=A+B', \ \ \ p={1\over 2}\tilde{u}C^{-1}u, 
\ \ \ p'={1\over 2}\tilde{v'}C^{-1}v' , \ \ \ q={1\over 2} \Bigg(
\tilde{u}C^{-1}v'+\tilde{v'}C^{-1}u\Bigg).
\end{equation}
Here we used the
familiar formula of the $3n$-dimensional Gaussian integration 
\begin{equation}
\int d{\bf x}\,\ {\rm e}^{-{1\over 2}\tilde{{\bf x}}A {\bf x}+
\tilde{{\bf T}}{\bf x}}=
\left({(2\pi)^{n} \over {\rm det} A }\right)^{3/2}
{\rm e}^{{1\over 2}\tilde{{\bf T}} A^{-1} {\bf T}},
\end{equation}
where $A$ is an $n\times n$ symmetric matrix and 
$\tilde{{\bf T}}=({\bf T}_1,\cdot\cdot\cdot,{\bf T}_n)$ is a row vector 
comprising three-dimensional vectors ${\bf T}_i$. 
Differentiating with respect to $\alpha$ and $\alpha'$, followed by 
$\alpha=\alpha'=0$, and integrating over the angles of ${\bf t}$ and 
${\bf t}'$ leads us to   
\begin{eqnarray}
& &\langle f_{KLM}(u,A,{\bf y})\vert f_{K'LM}(v,B,{\bf x})\rangle
\nonumber \\
&=&\frac{1}{B_{KL}B_{K'L}}
\left({(2\pi)^{N-1} \over {\rm det}C}\right)^{3/2}
\kappa!\kappa'! \sum_{n=0}^{{\rm min}(K,K')}
{{p^{K-n}{p'}^{K'-n}}q^{L+2n} \over 
({K-n})!
({K'-n})! (L+2n)!}B_{nL}.
\end{eqnarray}

\par\indent
The matrix element of the kinetic energy operator is calculated in a 
similar way. Expressing the intrinsic kinetic energy operator 
in terms of the set of coordinates $\bf y$ of the bra side, we obtain 
\begin{eqnarray}
& &\langle f_{KLM}(u,A,{\bf y})\vert 
\sum_{i=1}^{N-1} {{\bf p}_i^2\over 2 \mu_i}\vert
f_{K'LM}(v,B,{\bf x})\rangle=\langle f_{KLM}(u,A,{\bf y})\vert 
\sum_{i=1}^{N-1} {{\bf p}_i^2\over 2 \mu_i}\vert
f_{K'LM}(v',B',{\bf y})\rangle
\nonumber \\
&=&\frac{1}{B_{KL}B_{K'L}} \int\!\int d{\hat {\bf t}}d{\hat {\bf t}}'
Y_{LM}({\hat {\bf t}})^{*} Y_{LM}({\hat {\bf t}}') \Bigg( 
{d^{\kappa+\kappa'} \over d\alpha^{\kappa}
d\alpha'^{\kappa'}}\,\ \langle g(\alpha,{\bf t},u,A,{\bf y})\vert 
\nonumber \\
&\ &\times \Big[3{\rm Tr}\Lambda B'-\alpha'^2 \tilde{v'}\Lambda v'+
2\alpha'{\bf t}'\cdot (\tilde{v'}\Lambda B'{\bf y})-
\tilde{{\bf y}}B'\Lambda B'{\bf y}\Big]\,\ \vert 
g(\alpha',{\bf t}',v',B',{\bf y})\rangle 
\Bigg)_{\alpha=\alpha'=0 \atop t=t'=1}.
\end{eqnarray}

Here $\Lambda$ is an $(N-1)\times(N-1)$ diagonal matrix
\begin{eqnarray}
\Lambda=
\left(
\begin{array}{cccc}
 \hbar^2\over2\mu_1 & 0 & ...&  0       \\
 0  & \hbar^2\over2\mu_2 &    &  \vdots    \\
\vdots   &       &      &  \vdots         \\ 
 0  &...  & ...  & \hbar^2\over2\mu_{N-1} 
\end{array}
\right). 
\end{eqnarray}

\par\indent
The integration over $\bf y$ in Eq. (20) can be done by using the formula 
of Eq. (18) as follows:
\begin{eqnarray}
& &\int d{\bf x} \,\ \tilde{{\bf U}}{\bf x}{\rm e}^{-{1\over 2}
\tilde{{\bf x}}A {\bf x}+\tilde{{\bf T}}{\bf x}}=
\Bigg(\frac{d}{d\alpha}\int d{\bf x}\,\ {\rm e}^{-{1\over 2}\tilde{{\bf x}}A 
{\bf x}+(\tilde{{\bf T}}+
\alpha\tilde{{\bf U}}){\bf x}}\Bigg)_{\alpha=0} \nonumber \\
&=&\left({(2\pi)^{n} \over {\rm det} A }\right)^{3/2}
{1 \over 2}(\tilde{{\bf U}}A^{-1}{\bf T}+\tilde{{\bf T}}A^{-1}{\bf U})
{\rm e}^{{1\over 2}\tilde{{\bf T}} A^{-1} {\bf T}}, \\
& &\int d{\bf x} \,\ \tilde{{\bf x}}B {\bf x} {\rm e}^{-{1\over 2}
\tilde{{\bf x}}A 
{\bf x}+\tilde{{\bf T}}{\bf x}}=\Bigg(-2\frac{d}{d\alpha}
\int d{\bf x}\,\ {\rm e}^{-{1\over 2}\tilde{{\bf x}}(A+\alpha B) 
{\bf x}+\tilde{{\bf T}}{\bf x}}\Bigg)_{\alpha=0} \nonumber \\
&=&\left({(2\pi)^{n} \over {\rm det} A }\right)^{3/2}
(3{\rm Tr}A^{-1}B+\tilde{{\bf T}}A^{-1}BA^{-1}{\bf T})
{\rm e}^{{1\over 2}\tilde{{\bf T}} A^{-1} {\bf T}}.
\end{eqnarray}
Using Eqs. (22) and (23) in Eq. (20) leads us to 

\begin{eqnarray}
& &\langle f_{KLM}(u,A,{\bf y})\vert 
\sum_{i=1}^{N-1} {{\bf p}_i^2\over 2 \mu_i}\vert
f_{K'LM}(v,B,{\bf x})\rangle
\nonumber \\
&=&\frac{1}{B_{KL}B_{K'L}} \int\!\int d{\hat {\bf t}}d{\hat {\bf t}}'
Y_{LM}({\hat {\bf t}})^{*} Y_{LM}({\hat {\bf t}}') 
\Bigg( {d^{\kappa+\kappa'} \over d\alpha^{\kappa}d\alpha'^{\kappa'}}
\nonumber \\
&\ &\times \left({(2\pi)^{N-1} \over {\rm det}C}\right)^{3/2}
\Big[R+P\alpha^2+P'\alpha'^2+Q\alpha \alpha' {\bf t}\cdot{\bf t'}\Big]
{\rm e}^{p\alpha^2+p'\alpha'^2+q\alpha\alpha' {\bf t}\cdot{\bf t'}}
\Bigg)_{\alpha=\alpha'=0 \atop t=t'=1},
\end{eqnarray}
where 
\begin{eqnarray}
R&=&3{\rm Tr}\Lambda B'C^{-1}A,\ \ \ 
P=-\tilde{u}C^{-1}B'\Lambda B'C^{-1}u,\ \ \ 
P'=-\tilde{v'}C^{-1}A\Lambda AC^{-1}v', \nonumber \\
Q&=&\tilde{u}C^{-1}B'\Lambda AC^{-1}v' 
+\tilde{v'}C^{-1}A\Lambda B'C^{-1}u.
\end{eqnarray}
The matrix element of the kinetic energy operator is finally obtained as 
\begin{eqnarray}
& &\langle f_{KLM}(u,A,{\bf y})\vert 
\sum_{i=1}^{N-1} {{\bf p}_i^2\over 2 \mu_i}\vert
f_{K'LM}(v,B,{\bf x})\rangle
\nonumber \\
&=&\frac{1}{B_{KL}B_{K'L}}
\left({(2\pi)^{N-1} \over {\rm det}C}\right)^{3/2}
\kappa!\kappa'! \sum_{n=0}^{{\rm min}(K,K')} \nonumber \\
&\ &\times\Big[Rpp'q+(K-n)Pp'q+(K'-n)pP'q+(L+2n)pp'Q\Big] \nonumber \\
&\ &\times{{p^{K-n-1}{p'}^{K'-n-1}}q^{L+2n-1} \over ({K-n})!
({K'-n})! (L+2n)!}B_{nL}. 
\end{eqnarray}
The present formulation does not have any problem arising from the 
center-of-mass motion as discussed in \cite{KA91,KP93}. 

\par\indent
We will show the evaluation of the potential energy matrix element by 
assuming that the potential $V_{ij}$ is a function of the  
distance $\vert {\bf r}_i - {\bf r}_j \vert$ only. There is at least one set 
of coordinates, say, ${\bf z}$, in which $({\bf r}_i - {\bf r}_j)$ can be 
chosen to be ${\bf z}_1$. It is then convenient to calculate the matrix 
element of the potential by transforming both of the 
correlated Gaussians on bra and ket sides to those expressed in 
this set of coordinates:  
\begin{eqnarray}
& &\langle f_{KLM}(u,A,{\bf y})\vert V_{ij} \vert
f_{K'LM}(v,B,{\bf x})\rangle
=\langle f_{KLM}(u',A',{\bf z})\vert V(z_1) \vert
f_{K'LM}(v',B',{\bf z})\rangle
\nonumber \\
&=&\frac{1}{B_{KL}B_{K'L}} \int\!\int d{\hat {\bf t}}d{\hat {\bf t}}'
Y_{LM}({\hat {\bf t}})^{*} Y_{LM}({\hat {\bf t}}') \nonumber \\
&\ &\times \Bigg( {d^{\kappa+\kappa'} \over d\alpha^{\kappa}
d\alpha'^{\kappa'}}\,\ \langle g(\alpha,{\bf t},u',A',{\bf z})\vert 
V(z_1) \vert g(\alpha',{\bf t}',v',B',{\bf z})\rangle 
\Bigg)_{\alpha=\alpha'=0 \atop t=t'=1},
\end{eqnarray}
where $A'$ and $u'$ are defined by Eq. (8) with the matrix $T$ 
corresponding to the ${\bf y}\to{\bf z}$ transformation ${\bf y}=T
{\bf z}$, while $B'$ and 
$v'$ are obtained similarly by the matrix 
corresponding to the ${\bf x}\to{\bf z}$ transformation. 

\par\indent
To perform the integration over $\bf z$ in Eq. (27) we introduce the 
short-hand notation as follows:
\begin{eqnarray}
\tilde{u'}&=&(u_1',u_2',\cdot\cdot\cdot,u_{N-1}')=(u_1',\tilde{\omega}),
\ \ \ \ \ \ 
\tilde{v'}=(v_1',v_2',\cdot\cdot\cdot,v_{N-1}')=(v_1',\tilde{\chi}), 
\nonumber \\ 
&\,\ &A'+B'=
\left(
\begin{array}{cccc}
 c         & \gamma_1 & ...   &  \gamma_{N-2}    \\
 \gamma_1  &          &       &                  \\
 \vdots    &          & \Gamma  &                  \\ 
 \gamma_{N-2}&        &       &  
\end{array}
\right). 
\end{eqnarray}
By introducing $\tilde{\gamma}=(\gamma_1,\cdot\cdot\cdot, \gamma_{N-2})$ and 
$\tilde{{\bf z}}=
({\bf z}_1,\cdot\cdot\cdot,{\bf z}_{N-1})=({\bf z}_1,\tilde{{\bf w}})$, 
the {\bf z} integration of Eq. (27) becomes 

\begin{eqnarray}
& &\langle g(\alpha,{\bf t},u',A',{\bf z})\vert 
V(z_1) \vert g(\alpha',{\bf t}',v',B',{\bf z})\rangle = \int\!\int 
d{{\bf z}_1}d{\bf w} \,\ V(z_1) \nonumber \\
&\ &\times {\rm exp}\Big(
-{1 \over 2}c{z_1}^2+(\alpha u_1'{\bf t}+\alpha 'v_1'{\bf t}'
)\cdot{\bf z}_1-{1 \over 2}\tilde{{\bf w}}\Gamma{\bf w}+
\Big[\alpha {\bf t}\cdot (\tilde{\omega}{\bf w})
+\alpha '{\bf t}'\cdot (\tilde{\chi}{\bf w})-{\bf z}_1 \cdot (\tilde{\gamma}
{\bf w})\Big] \Big) \nonumber \\
&\ &=\left({(2\pi)^{N-2} \over {\rm det}\Gamma}\right)^{3/2}
{\rm e}^{p_{v}\alpha^2+p_{v}'\alpha'^2+q_{v}\alpha\alpha' {\bf t}
\cdot{\bf t'}} \int d{{\bf z}_1}\,\ V(z_1)
{\rm e}^{-{1 \over 2}(c-\tilde{\gamma} \Gamma ^{-1}
\gamma)z_1^{2}+(\lambda \alpha {\bf t}+\lambda '\alpha '{\bf t}')
\cdot{\bf z}_1} \\
&\ &=\left({(2\pi)^{N-2} \over {\rm det}\Gamma}\right)^{3/2}
{\rm e}^{p_{v}\alpha^2+p_{v}'\alpha'^2+q_{v}\alpha\alpha' {\bf t}
\cdot{\bf t'}} 4\pi \int _{0}^{\infty} d{z_1}\,\ z_1^{2}V(z_1) 
{\rm e}^{-{1 \over 2}
(c-\tilde{\gamma} \Gamma ^{-1}\gamma)z_1^{2}}\,\ i_{0}(\vert 
\lambda \alpha {\bf t}
+\lambda '\alpha '{\bf t}' \vert z_{1}) \nonumber.
\end{eqnarray}
where 
\begin{eqnarray}
& &p_v={1\over 2}\tilde{\omega} \Gamma^{-1}\omega, 
\ \ \ p_{v}'={1\over 2}\tilde{\chi} \Gamma^{-1}\chi , \ \ \ q_v=
{1\over 2} \Bigg(\tilde{\omega} \Gamma^{-1}\chi+\tilde{\chi} \Gamma^{-1}
\omega\Bigg). 
\nonumber \\
& &\lambda=u_{1}'-{1 \over 2}\Bigg(\tilde{\gamma} \Gamma^{-1}\omega+
\tilde{\omega} \Gamma^{-1}\gamma\Bigg), \ \ \ 
\lambda '=v_{1}'-{1 \over 2}\Bigg(\tilde{\gamma} \Gamma^{-1}\chi+
\tilde{\chi} \Gamma^{-1}\gamma\Bigg),
\end{eqnarray}
and $i_{0}(x)={\rm sinh}x/x$. 
Substituting Eq. (29) into Eq. (27), followed by a power series expansion 
of $i_{0}(x)$, and doing the needed operation leads us to 

\begin{eqnarray}
& &\langle f_{KLM}(u,A,{\bf y})\vert V(z_1) \vert
f_{K'LM}(v,B,{\bf x})\rangle \nonumber \\ 
&=&\frac{1}{B_{KL}B_{K'L}}\left({(2\pi)^{N-2} \over {\rm det}\Gamma}
\right)^{3/2}\kappa !\kappa '! \sum_{m=0}^{K+K'+L}\sum_{i=0}^{m}
\sum_{j=0}^{m-i}\sum_{n={\rm max}(0,{m-i-j-L \over 2})}^
{{\rm min}(K-i,K'-j)} I(2m+2,c-\gamma \Gamma^{-1}
\gamma) \nonumber \\
&\ & \times {2^{m-i-j}m!\lambda^{m+i-j}\lambda '^{m-i+j}p_{v}^{K-i-n}
p_{v}'^{K'-j-n}q_{v}^{L+2n-m+i+j} \over (2m+1)!i!j!(m-i-j)!(K-i-n)!
(K'-j-n)!(L+2n-m+i+j)!}B_{nL},
\end{eqnarray}
with 
\begin{equation}
I(n,a)=4\pi\int _{0}^{\infty} d{z_1}\,\ V(z_1)z_1^{n} 
{\rm e}^{-{1 \over 2}az_1^{2}}.
\end{equation}
The integral of Eq. (32) becomes elementary for the Coulomb potential. 
The calculation of the mean distance, the root mean square distance or 
the mean inverse distance between the particles can easily be done by 
putting $V(z)=z$, $z^2$ or $1/z$ in the above integral. 

\par\indent
The calculation of the matrix elements described above is much simpler  
than the case where the function $\theta_{LM_L}({\bf x})$ 
is decomposed into partial waves of the relative coordinates as in Eq. (3). 
In fact, in that latter
case one has to integrate over the angles of the relative 
coordinates and one has to cope with the angular momentum algebra. 
We note, however, that the calculation of the matrix element of the latter 
type poses no problem if the function $\theta_{LM_L}({\bf x})$ of Eq. (3) 
is expressed as a linear combination of the terms of Eq. (4) with 
appropriate $u$-vectors. It is appealing that the present formalism 
does not require any modification with respect to an increasing $N$. 

\par\indent
All the matrix elements can be given in a closed analytic
form and the numerical evaluation of the matrix elements as a function
of the nonlinear parameters is therefore straightforward. 
The values, $\kappa=2K\!\!+\!\!L$ and $\kappa'=2K'\!\!+\!\!L$, are usually 
small in practical cases and the sum in Eqs. (19), (26), and (31) is 
limited to just a few terms.

\par\indent
The partial derivative of matrix elements with respect to  
variational parameters may sometimes be useful when one searches for 
an optimal set of parameters. Since the dependence on the parameters 
is explicitly given for the matrix elements of the correlated Gaussians, 
it would not be difficult to 
derive the expression for the derivative. As stated in the previous 
section, the $u$-vector is considered a variational parameter which  
defines the most suitable global vector to describe the rotational motion. 
The calculation of the derivative of the matrix elements with respect 
to $u_i$ is particularly simple because of the simple structure of the 
$u$-dependence. It can also be calculated by using an equation analogous 
to Eq. (14) because the derivative of $\eta_{KLM}$ with respect to 
$u_{i}$ is expressed as a tensor product of two $\eta$'s as follows: 

\begin{eqnarray}
\frac{\partial}{\partial u_{i}}\eta_{KLM}(u,{\bf x})&=&(2K+2L+1)
\sqrt{{L\over 2L+1}}\Big[\eta_{KL-1}(u,{\bf x}){\bf x}_{i}
\Big]_{LM} \nonumber \\
&-&2K\sqrt{{L+1\over 2L+1}}\Big[\eta_{K-1L+1}(u,{\bf x})
{\bf x}_{i}\Big]_{LM}.
\end{eqnarray}
Note that ${\bf x}_i$ is equal to $\sqrt{4\pi/3}\eta_{01}(u',{\bf x})$ 
with $u'$ being defined by $u'_{j}=\delta_{ij}$.
 
\section{Specific examples: Coulomb three-body system} 

\par\indent
This section is devoted to show the bound-state solutions of Coulomb 
three-body 
problems by using the method described above. To test the method we 
consider the positronium negative ion $e^{-}e^{-}e^{+}$, the muonic 
molecules of $tt\mu^{-}$ and $td\mu^{-}$, and the helium atom 
$\alpha e^{-}e^{-}$. These systems are 
chosen because they cover a wide range of mass ratio of the constituent 
particles and include some rotational bound states:  
The positronium negative ion consists of particles of 
identical mass, which makes it difficult to treat the system in an adiabatic 
approach, the mass of $t$ or $d$ in the muonic molecule is a few ten times 
heavier than the mass of $\mu$, and in the helium atom the mass of the 
heavy particle is several thousand times larger than the $e^-$ mass, 
which is the realm of the adiabatic approach. One can in principle 
consider other systems with the method, but there are already 
highly accurate calculations available for the above systems. We assume 
that the orbital wave function 
is antisymmetric with respect to the interchange of the two identical  
particles except for the case of the spin-singlet states of the helium atom. 

\par\indent
The trial wave function is chosen to be a linear combination of the 
correlated Gaussian of type (6). Without loss of generality the vector $u$ 
can be set to satisfy $u_{1}^2+u_{2}^2=1$. Each basis function thus 
contains at most 
four nonlinear parameters, three of which come from the matrix $A$. 
To assure the positive definiteness of the correlated Gaussians, 
$A$ can in general be expressed as $A =\tilde{G}DG$ \cite{KP93}, 
where $G$ is an 
$(N\!-\!1)\!\times\!(N\!-\!1)$ orthogonal matrix containing 
$(N\!-\!1)(N\!-\!2)/2$ 
parameters and $D$ is a diagonal 
matrix, $D_{ij}=d_{i}
\delta_{ij}$, including $(N\!-\!1)$ positive parameters $d_i$.  Although no 
restriction on the parameters of the matrix $G$ 
is in principle necessary, it is advisable to avoid too many variables  
if possible. The most naive choice would be to take $G$ as a 
unit matrix, which is equivalent to using only a single set of 
coordinates $\bf x$, and then to try to reach 
convergence by including successively higher partial waves implied by an 
appropriate choice of $K$ and $u$ values. Many examples 
show \cite{prca,kamimura,SVAO}, however, that this type of single channel 
calculations does not work well especially in those systems where the 
adiabatic approximation of the motion is questionable because the various 
types of correlated motion become important. Instead the matrix $A$ may  
be chosen as $A=\widetilde{T^{-1}}DT^{-1}$, where $T$ is such a matrix 
that connects the coordinates $\bf x$ to an arbitrary set of relative  
coordinates $\bf y$, ${\bf x}=T{\bf y}$. In this case 
$\tilde{{\bf x}}A{\bf x}$ becomes 
$\tilde{{\bf y}}D{\bf y}=\sum_{i=1}^{N-1}d_iy_i^{2}$. The Gaussians 
with this choice of $A$ together with the angular part of Eq. (3) 
have recently been shown to give a precise solution for 
few-nucleon systems with realistic nucleon-nucleon potentials 
including noncentral components \cite{vos}. 
Since our interest is to test the utility of Eq. (6), we restricted 
the choice of $A$ and a maximum value of $K$ to the following three 
simple cases:

\begin{list}{(i)}{}
\item  $K_{\rm max}=0$ and $A=\widetilde{T^{-1}}DT^{-1}$, where $T$ 
represents a special matrix connecting to three sets of 
relative coordinates (the so-called rearrangement channels) and $D$ 
is diagonal. 
\end{list}
\begin{list}{(ii)}{}
\item  $K_{\rm max}=1$ and the choice of $A$ is the same as in case 
(i). 
\end{list}
\begin{list}{(iii)}{}
\item  $K_{\rm max}=0$ but $A$ is an arbitrary positive-definite, 
symmetric matrix.
\end{list} 
In case (iii) the matrix $A$ can be parametrized as $A =\tilde{G}DG$ 
with an orthogonal matrix $G$. As was already 
mentioned in the previous section, the angular part with $K=0$ describes 
the stretched configuration and therefore allows such limited angular 
correlations in case (i). The case (ii) is an extension of this case to 
include the non-stretched coupling. The angular correlation between the 
particles is taken into account in case (iii) by the cross term of the 
exponential part of the correlated Gaussian. 

\par\indent
Our primary purpose is to investigate if our basis functions with the 
new angular part can universally yield satisfactory 
results of the same quality independently of the system. 
Results of the calculation presented below are thus limited up to the 
maximum basis dimension ${\cal K}=200$ for all the systems. With a larger 
basis size one can definitely obtain much more accurate results but this is 
beyond our purpose. Even with ${\cal K}=200$ dimension 
there are a large number of nonlinear parameters. A complete optimization 
is in principle superior to any other method but it is neither 
practical nor possible at present. Instead we select the basis elements 
in the trial function stepwise according to the stochastic variational 
method (SVM) \cite{prca,SVM,VSL,SVAO,vos}. The SVM 
attempts to select the most appropriate basis elements in a trial and error 
procedure: Various randomly generated candidates for the basis element are 
tested and the usefulness of these states are judged by their contribution 
to the energy of the system. The SVM has proved to provide a precise  
solution for various few-body systems with a reasonable computational 
effort. The refining procedure employed in \cite{vos} to tune the nonlinear 
parameters was effective to reach the solution of high quality for three- 
and four-nucleon systems interacting with realistic nuclear potentials and 
was successfully used in the present paper as well. The basis selection 
in the SVM can be done by exploiting the special form of the Hamiltonian 
matrix as shown in Ref. \cite{prca} and it does not carry heavy 
computational loads. There are several sophisticated optimization 
strategies in quantum chemistry \cite{temper,cencek,bukowski} including,  
for example, the random tempering \cite{temper,alexander}. 
The main difference between the SVM and other random optimization methods 
such as the random tempering is that the SVM employs the step-by-step 
procedure to build up a basis set, optimizing the nonlinear parameters of 
the basis states with respect to each other. The advantage of using 
the SVM here is that it seems to be quite suitable to find the optimal 
$u$ vector. 

\par\indent
The results for the positronium negative ion are compared to other 
calculations \cite{ho,bhatia83,yeremin,krivec,ps-,ho93} in Table II. The 
energy obtained at the dimension $\cal K$ is also shown. The trial function 
of case (i) does not produce good result, which is not surprising because 
the 
interparticle correlation is poorly represented in this case. The result 
of case (ii) shows a significant improvement over the case (i), 
confirming the importance 
of the polynomial part of type (4). The energy and the root mean 
square radius in this case are slightly better than our previous calculation 
\cite{prca}. This improvement is due to the refinement in 
the optimization of the nonlinear parameters, which were not employed 
before.   
A full correlated Gaussian basis of case (iii), though no polynomial part 
is employed, gives 
even better results which are comparable to the results obtained in the 
similar basis dimension by using the generalized 
Hylleraas-type wave functions \cite{ho,bhatia83}. Compared to the 
correlation-function hyperspherical harmonic (CFHH) method \cite{krivec}, 
our correlated 
Gaussian basis seems to give a better solution. The table also 
lists the calculated average distances between the two electrons and 
between the positron and an electron. They are in reasonable agreement with 
the most precise values obtained by using the correlated exponential (CE) 
functions \cite{ps-} or the Hylleraas-type functions \cite{ho93}.  

Table III lists the results for the lowest $L=0-3$ states of the $tt\mu$ 
molecule. This system is analogous to the negative positronium ion 
though the mass ratio of the constituent particles is considerably 
different. The results with the cases of (ii) and 
(iii) are shown and the quality of them is similar to what is 
mentioned in the case of the negative positronium ion. Our calculation 
for the $S$ and $P$ states reproduces the first seven digits of Ref. 
\cite{alexander} which uses  the correlated Slater-type geminals or 
interparticle CE functions. For the 
$D$ state our result is in good agreement with the extensive calculation 
of \cite{frolov94} using ${\cal K}=2250$ basis functions similar to 
\cite{alexander}. 
Both of them employ the bipolar harmonics of the stretched coupling 
to describe the rotational motion. The calculation of case (iii) uses 
$K_{\rm max}=0$ so that the basis function in this case also employs 
only the stretched coupling. The 
fact that we have got rather accurate results in a small basis suggests that 
the usage of Eq. (6), particularly its angular part is very useful. We 
confirm that the $F$ state is bound, in agreement with the other calculation 
\cite{vinit}, but have found no bound $G$ state. 

\par\indent
We next show in Table IV the results of calculation for the $dt\mu$ 
molecule. The basis functions of case (iii) again give better 
energies than those of case (ii). In fact they reproduce the first six 
digits of the most precise variational calculations 
\cite{kamimura,alexander,frolov94} for $L=0-2$ states. 
The Gaussian basis similar to case (ii) is employed in the GBCRC 
calculation of \cite{kamimura}, where the angular part is, however, 
represented by the successive coupling of type (3). The fact that the 
$D$ state energy of our calculation with ${\cal K}=200$ becomes slightly 
lower than that of the GBCRC calculation with ${\cal K}=1566$ confirms 
that a careful optimization of the nonlinear parameters is very important 
and moreover that the angular function of Eq. (4) is really useful. 
In Ref. \cite{alexander} 
the optimized energy at smaller $\cal K$ values is given for the $S$ and $P$ 
states. Our energies are slightly lower than theirs at the same dimension, 
which indicates the effectiveness of the basis optimization of the SVM.

\par\indent
Finally we present in Table V the energies of the helium atom which 
are obtained with cases (ii) and (iii). We used the finite mass of the 
$\alpha$-particle, $m_{\alpha}=7294.2618241m_e$. The basis functions of 
case (iii) give lower energies than those of (ii) for the $S$ and $P$ 
states but produce considerably higher energies 
for the states with higher $L$ values. This suggests that 
the explicit introduction of the non-stretched coupling is more favorable 
to describe in our formulation the rotational motion of highly asymmetric 
states. In the table the energies of the $S$ and $P$ states are compared to 
the result of Ref. \cite{frolov86} where the bound states of the atomic 
and mesomolecular systems were studied up to $L=2$ by using the  
CE functions of the interparticle distances. The angular part 
is expressed by the bipolar harmonics. The agreement is fair. 
To include the comparison 
with the combined configuration interaction Hylleraas method 
calculation \cite{sims}, we repeated the calculation for the $D$ and 
$F$ states assuming the infinite mass for the $\alpha$-particle. 
We see that our basis functions reproduce the energies of both states 
up to the first six-seven digits. The advantage of our method is that no 
special care is needed to treat states with high angular momenta. 

\section{Summary}
We have presented a formulation for the correlated Gaussian basis which 
can describe the orbital motion with an arbitrary angular momentum. 
Instead of the well-known 
successive couplings of the spherical harmonics to the total orbital 
angular momentum the angular part of the basis 
functions in this formulation is uniquely represented by a single spherical 
harmonic of a global vector which is a linear combination of the 
relative coordinates. We have shown that 
this type of the correlated Gaussians is simply derived from the 
generating function which is invariant with respect to the transformation 
of the relative coordinates. This property makes it much simpler to 
calculate the matrix elements and in fact the formulas 
for the matrix elements have been derived in a 
compact form for a general system of $N$-particles interacting with the 
central forces. They can be applied universally to diverse systems 
independently of the number of particles and of the total angular momentum. 
The extension to include the noncentral forces is straightforward. It 
should be noted that the present formulation for the angular part can  
be applied to other orbital functions as well as the correlated Gaussian. 

The formulation has been applied to obtain the bound states of various 
Coulomb three-body systems which cover a wide range of mass ratio, 
namely, $e^-e^-e^+,\, tt\mu,\, td\mu$, and $\alpha e^-e^-$. 
The basis functions have been set up stepwise in a trial and error procedure 
using the stochastic variational method. The coefficients defining 
the global vector can be varied continuously to minimize the energy. 
The nonlinear parameters 
of the correlated Gaussians are parametrized by three different options. 
One of the options is to include only the so-called rearrangement channels 
with the angular functions which describe the non-stretched coupling. This 
model has already given fairly good results. Compared to the restricted 
case in which only the stretched coupling is included, this option 
presents much better results. The other option 
is to use the full matrices to include the correlation between the 
particles. This model, even without use of the polynomial part, has been 
better than the former model in most cases but the high $L$ states 
of the helium atom and, with the basis size of 200, reproduced 
the first six$-$seven digits of the total binding energies of almost all the 
systems mentioned above. We have tested the bound 
states up to $L=4$ and found that no difficulty arises from the 
treatment of high angular momentum states in our formulation. We have 
confirmed that the correlated Gaussian 
of type (6) works very nicely as a basis function for bound-state solution 
for diverse systems. Although we suggested the method to treat the 
unnatural parity case, a more unified representation for both natural and 
unnatural parity cases should be developed. 

Finally we summarize some merits of our method in the following.
\begin{list}{(i)}{}
\item  No partial wave expansion is needed, and thus no problems in 
angular momentum coupling arise. 
\end{list}
\begin{list}{(ii)}{}
\item  Universality of the scheme. One needs to introduce no change, 
for example, when treating a larger system of $N$-particles or describing 
states with high angular momentum. There is no center-of-mass motion 
problem.  
\end{list}
\begin{list}{(iii)}{}
\item  Invariance with respect to the coordinate transformation. The 
form of basis states is kept invariant under the transformation. It is 
easy to construct the symmetry-adapted basis states. 
\end{list}
\begin{list}{(iv)}{}
\item  Fully analytical calculational scheme. The matrix elements are 
evaluated as simply as those for $L=0$. The dependence of the matrix 
elements on the nonlinear parameters has simple structure. 
\end{list}  
 
\bigskip
\par\indent
This work was supported by Grants-in-Aid for Scientific Research 
(No. 05243102 and No. 06640381) and for International Scientific Research 
(Joint Research) (No. 08044065) of the Ministry of Education, Science 
and Culture (Japan) and by OTKA grant No. T17298 (Hungary). The authors 
are grateful for the use of RIKEN's computer facility which 
made possible most of the calculations. 
Thanks are also  due to the support of both Japan Society for the Promotion 
of Science and Hungarian Academy of Sciences. 

\section*{Appendix}
We prove by induction that the vector-coupled function of two 
solid spherical harmonics 
\begin{equation}
\Big[{\cal Y}_{l_1}
({\bf x}_1) {\cal Y}_{l_2}({\bf x}_2)\Big]_{LM} \ \ \ 
{\rm with} \ \ \ (-1)^{l_1+l_2}=(-1)^L
\end{equation}
can be expressed in terms of a linear combination of terms 
\begin{equation}
\vert {\bf x}_1\vert^{2p_1}\vert {\bf x}_2\vert^{2p_2}\vert {\bf v} 
\vert^{2q+L}Y_{LM}(\hat{\bf v}),
\end{equation}
where the degree of the function is given by $2p_1+2p_2+2q+L=l_1+l_2$,
and the vector ${\bf v}$ of each term is given by 
${\bf v}=u_1{\bf x}_1+u_2{\bf x}_2$ with appropriate 
coefficients $u_1$ and $u_2$.  

First we prove that the statement is true for a special case of 
$l_1+l_2=L$, namely for the lowest order terms for a given $L$. As 
Eq. (5) shows, $\vert {\bf v}\vert^{L}
Y_{LM}(\hat{\bf v})$ consists of $L+1$ terms of $\Big[{\cal Y}_{l}
({\bf x}_1) {\cal Y}_{L-l}({\bf x}_2)\Big]_{LM}$ ($l=0,1,\cdot\cdot\cdot,
L$), each multiplied by 
$u_{1}^{l} u_{2}^{L-l}$. By using $L+1$ mutually different $u_1$ values, 
$(u_1^{(1)}, 
u_1^{(2)}, \cdot\cdot\cdot, u_1^{(L+1)})$, in Eq. (5) and keeping 
$u_2$ an arbitrary constant, it is possible to pick up a particular term 
$\Big[{\cal Y}_{l}({\bf x}_1) {\cal Y}_{L-l}({\bf x}_2)\Big]_{LM}$ 
through a linear combination of 
$\sum_{\alpha=1}^{L+1} c_{\alpha}\vert {\bf v}_{\alpha}\vert^{L}
Y_{LM}(\hat{\bf v}_{\alpha})$ with ${\bf v}_{\alpha}=u_1^{(\alpha)}{\bf x}_1
+u_2{\bf x}_2$, where $c_{\alpha}$ should satisfy the following equation
\begin{equation}
\sum_{\alpha =1}^{L+1} (u_1^{(\alpha)})^{m} c_{\alpha} = \delta_{ml}, \ \  
\ {\rm for} \ \ \ m=0,1,\cdot\cdot\cdot,L.
\end{equation}
The solution of the above linear equation for $c_{\alpha}$ 
is given by Vandermonde's determinant. Thus our assertion is proved. 

Next we assume that the statement holds for all the cases of 
$l_1+l_2 \le 2(K-1)+L$. We note that all the terms of   
$\Big[{\cal Y}_{K+l}({\bf x}_1) {\cal Y}_{K+L-l}({\bf x}_2)\Big]_{LM}$ 
($l=0,1,\cdot\cdot\cdot,L$) with degree $2K+L$ appear in the expansion of 
Eq. (5) together with the respective coefficients $u_1^{K+l} u_2^{K+L-l}$. 
The rest of the terms in the expansion can, by assumption, be expressed in 
the form of Eq. (35) because the vector-coupled part of each term 
has smaller degree 
than $2K+L$. Using $L+1$ different $u_1$ values similarly to the above 
case of Eq. (36) enables us to express each of the terms, 
$\Big[{\cal Y}_{K+l}({\bf x}_1) {\cal Y}_{K+L-l}({\bf x}_2)\Big]_{LM}$, 
in terms of a linear combination of the functions of Eq. (35). 
This completes the proof.

\begin{table}

\caption{Possible sets of partial waves contained in a single spherical 
harmonic with low $L$ and $K$ values for a three-body system. See Eq. (5).  
Listed below are only the cases of $2k_1+l_1 \le 2k_2+l_2$.}

\begin{tabular}{p{1cm}p{1cm}p{1cm}p{1cm}p{1cm}p{1cm}}\hline
$L$ & $K$ & $k_1$ & $l_1$ & $k_2$ & $l_2$ \\
\hline
 0  &  0  &  0  &  0  &  0  &  0  \\
\cline{2-6} 
    &  1  &  0  &  0  &  1  &  0  \\
    &     &  0  &  1  &  0  &  1  \\
\cline{2-6} 
    &  2  &  0  &  0  &  2  &  0  \\
    &     &  0  &  1  &  1  &  1  \\
    &     &  0  &  2  &  0  &  2  \\
    &     &  1  &  0  &  1  &  0  \\
\hline
 1  &  0  &  0  &  0  &  0  &  1  \\
\cline{2-6} 
    &  1  &  0  &  0  &  1  &  1  \\
    &     &  0  &  1  &  0  &  2  \\
    &     &  0  &  1  &  1  &  0  \\
\cline{2-6} 
    &  2  &  0  &  0  &  2  &  1  \\
    &     &  0  &  1  &  1  &  2  \\
    &     &  0  &  1  &  2  &  0  \\
    &     &  0  &  2  &  0  &  3  \\
    &     &  0  &  2  &  1  &  1  \\
    &     &  1  &  0  &  1  &  1  \\
\hline
 2  &  0  &  0  &  0  &  0  &  2  \\
    &     &  0  &  1  &  0  &  1  \\
\cline{2-6} 
    &  1  &  0  &  0  &  1  &  2  \\
    &     &  0  &  1  &  0  &  3  \\
    &     &  0  &  1  &  1  &  1  \\
    &     &  0  &  2  &  0  &  2  \\
    &     &  0  &  2  &  1  &  0  \\
    &     &  1  &  0  &  0  &  2  \\
\cline{2-6} 
    &  2  &  0  &  0  &  2  &  2  \\
    &     &  0  &  1  &  1  &  3  \\
    &     &  0  &  1  &  2  &  1  \\
    &     &  0  &  2  &  0  &  4  \\
    &     &  0  &  2  &  1  &  2  \\
    &     &  0  &  2  &  2  &  0  \\
    &     &  1  &  0  &  1  &  2  \\
    &     &  0  &  3  &  0  &  3  \\
    &     &  0  &  3  &  1  &  1  \\
    &     &  1  &  1  &  0  &  3  \\
    &     &  1  &  1  &  1  &  1  \\
\hline
 3  &  0  &  0  &  0  &  0  &  3  \\
    &     &  0  &  1  &  0  &  2  \\
\cline{2-6} 
    &  1  &  0  &  0  &  1  &  3  \\
    &     &  0  &  1  &  0  &  4  \\
    &     &  0  &  1  &  1  &  2  \\
    &     &  0  &  2  &  0  &  3  \\
    &     &  0  &  2  &  1  &  1  \\
    &     &  1  &  0  &  0  &  3  \\
\cline{2-6} 
    &  2  &  0  &  0  &  2  &  3  \\
    &     &  0  &  1  &  1  &  4  \\
    &     &  0  &  1  &  2  &  2  \\
    &     &  0  &  2  &  0  &  5  \\
    &     &  0  &  2  &  1  &  3  \\
    &     &  0  &  2  &  2  &  1  \\
    &     &  1  &  0  &  1  &  3  \\
    &     &  0  &  3  &  0  &  4  \\
    &     &  0  &  3  &  1  &  2  \\
    &     &  0  &  3  &  2  &  0  \\
    &     &  1  &  1  &  0  &  4  \\
    &     &  1  &  1  &  1  &  2  \\
\hline
\end{tabular}

\clearpage

\caption{The total binding energy and the distance of the 
$e^{-}e^{-}e^{+}$
 system. ${\cal K}$ is the number of basis functions used.
 See text for the cases (i), (ii), and (iii) of basis functions.
 Atomic units are used.}

\begin{tabular}{crllllll}\hline
&&&\multicolumn{3}{c}{$<r^{2}>^{1/2}$}&\multicolumn{2}{c}{$<r>$}\\
\cline{4-6}\cline{7-8}
\multicolumn{1}{c}{basis function}&\multicolumn{1}{r}{${\cal K}$}&
\multicolumn{1}{l}{energy}&\multicolumn{1}{c}{radius}&
\multicolumn{1}{l}{$e^-e^+$}&\multicolumn{1}{l}{$e^-e^-$}&
\multicolumn{1}{l}{$e^-e^+$}&\multicolumn{1}{l}{$e^-e^-$}\\
\hline
\\
(i)    & 50&0.26186974&&&\\
       &100&0.26188326&4.596&6.968&9.644&5.494&8.539\\
       &150&0.26188401&&&\\
       &200&0.26188445&4.597&6.96882&9.64479&5.49475&8.54000\\
\\
(ii)   & 50&0.26199779&&&\\
       &100&0.26200455&4.594&6.958&9.652&5.489&8.548\\
       &150&0.26200489&&\\
       &200&0.26200494&4.595&6.95835&9.65285&5.48963&8.54856\\
\\
(iii)  & 50&0.26199953&&&\\
       &100&0.26200465&4.594&6.958&9.652&5.489&8.548\\
       &150&0.26200491&&&\\
       &200&0.26200504&4.595&6.95812&9.65254&5.48957&8.54846\\
\\\hline
\\
Hylleraas\cite{ho} &125&0.262004895 &&6.957&9.650&5.4891&8.5476\\
Hylleraas\cite{bhatia83} &220&0.2620050565&\\
CFHH\cite{krivec} &225&0.262004673&&6.956&9.650&5.48881&8.54699\\
CE\cite{ps-}         &800&0.2620050702319&&6.95837&9.65291&5.4896332525
                                                          &8.5485806553\\
Hylleraas\cite{ho93} &744&0.2620050702328&&6.95837&9.65291&5.489633252
                                                          &8.548580655\\
\\\hline
\end{tabular}

\clearpage

\caption{The total binding energies and the rms distances of
 the lowest $S$, $P$, $D$, and $F$ states of the $tt\mu$ system.
 The mass set used is $m_t=5496.918 m_e$, and $m_{\mu}=206.7686 m_e$.
 Atomic units are used. }

\begin{tabular}{ccrllll}\hline

&&&&\multicolumn{3}{c}{$<r^{2}>^{1/2}\times10^{3}$}\\
\cline{5-7}
$L$&\multicolumn{1}{c}{basis function}&{${\cal K}$}&
\multicolumn{1}{l}{energy}&\multicolumn{1}{l}{radius}&
\multicolumn{1}{l}{$t\mu$}&\multicolumn{1}{l}{$tt$}\\
\hline
\\
$S$ &(ii)        &200&112.97253&7.444&11.15&13.39\\ 
  &(iii)       &200&112.97300&7.444&11.15&13.39\\
  & CE\cite{alexander} &500&112.9730179&\\
\\
$P$& (ii)        &200&110.26189&7.933&11.82&14.46\\ 
  & (iii)       &200&110.26210&7.933&11.82&14.46\\
  & CE\cite{alexander} &500&110.2621165&&&\\
\\
$D$& (ii)        &200&105.98288&8.919&13.18&16.59\\ 
  & (iii)       &200&105.98301&8.919&13.18&16.59\\
\\
  & (iii)$^{a}$ &200&105.98292&8.919&13.18&16.59\\
  & CE\cite{frolov94}$^{a}$ &2250&105.982930&&&\\
\\
$F$& (ii)         &200&101.43093&10.71&15.67&20.39\\ 
  & (iii)       &200&101.43105&10.71&15.67&20.38\\
\\
  & (iii)$^{b}$ &200&101.43131&10.71&15.67&20.39\\
  & Adiabatic \cite{vinit}$^{b}$ &   &101.43&&&\\
\\
\hline
\end{tabular}
$^a \;m_t=5496.92158m_e, m_{\mu}=206.768262m_e, 2R_{\infty}=27.2113961$eV.
\newline
$^b \;m_t=5496.918m_e, m_{\mu}=206.769m_e, R_{\infty}=13.6058$eV.

\clearpage

\caption{The total binding energies and the rms distances of the lowest
 $S$, $P$, and $D$ states of the $td\mu$ system.
 The mass set used is $m_t=5496.918m_e,\,m_d=3670.481m_e$, and 
$m_{\mu}=206.7686m_e$. Atomic units are used.}

\begin{tabular}{lcrlllll}\hline
&&&&\multicolumn{4}{c}{$<r^{2}>^{1/2}\times10^{3}$}\\
\cline{5-8}
$L$&\multicolumn{1}{c}{basis function}&\multicolumn{1}{r}{${\cal K}$}
&\multicolumn{1}{l}{energy}
&\multicolumn{1}{l}{radius}&\multicolumn{1}{l}{$d\mu$}
&\multicolumn{1}{l}{$t\mu$}&\multicolumn{1}{l}{$td$}\\
\hline
\\
$S$&(ii)   &100&111.36357&7.774&11.73&11.24&13.92\\
 &         &200&111.36398&7.774&11.73&11.24&13.92\\
\\
 &(iii)    &100&111.36363&7.774&11.73&11.24&13.92\\
 &         &200&111.36444&7.774&11.73&11.24&13.92\\
\\
 &\multicolumn{1}{c}{GBCRC\cite{kamimura}} &1442&111.364507&&&&\\
 &\multicolumn{1}{c}{CE\cite{alexander}} &1400&111.364511474&&&&\\
\\
\hline
\\
$P$&(ii)   &100&108.17803&8.417&12.68&12.03&15.31\\
 &         &200&108.17914&8.417&12.68&12.03&15.31\\
\\
 &(iii)    &100&108.17820&8.416&12.68&12.03&15.31\\
 &         &200&108.17940&8.417&12.68&12.03&15.31\\
\\
 &\multicolumn{1}{c}{CE\cite{alexander}}&1800&108.1795424&&&\\
\\
 &(iii)$^a$&200&108.17923&8.417&12.68&12.03&15.31\\
 &\multicolumn{1}{c}{GBCRC\cite{kamimura}$^a$} &2662&108.179385&&&\\

\\
\hline
\\
$D$&(ii)   &100&103.40632&9.766&14.81&13.61&18.19\\
 &         &200&103.40824&9.769&14.81&13.61&18.19\\
\\
 &(iii)    &100&103.40733&9.766&14.81&13.61&18.19\\
 &         &200&103.40849&9.769&14.81&13.61&18.19\\
\\
 &\multicolumn{1}{c}{GBCRC\cite{kamimura}} &1566&103.408481&\\
\\
\hline
\end{tabular}
$^a \;m_t=5496.92158m_e, m_d=3670.483014m_e, m_{\mu}=206.768262m_e, 
R_{\infty}=13.6056981$eV.

\clearpage

\caption{The total binding energies and the rms distances of the lowest
 $S$, $P$, $D$, $F$, and $G$ states of the helium atom $(\alpha \; e^- e^-)$.
 The correlated Gaussians of cases (ii) and (iii) are
 used for basis functions. The $\alpha$-particle mass is
 $m_{\alpha}=7294.2618241m_e$. Atomic units are used.}

\begin{tabular}{lcrlrrr}\hline
&&&&\multicolumn{3}{c}{$<r^{2}>^{1/2}$}\\
\cline{5-7}
$L$&\multicolumn{1}{r}{basis function}&\multicolumn{1}{r}{${\cal K}$}
&\multicolumn{1}{l}{energy}
&\multicolumn{1}{c}{radius}&\multicolumn{1}{c}{$e^-\alpha$}
&\multicolumn{1}{c}{$e^-e^-$}\\
\hline
\\
$1^1S$ & (ii)          &200&2.9033033&0.8920&1.093&1.587\\
       & (iii)         &200&2.9033041&0.8920&1.093&1.587\\
       & CE\cite{frolov86} &300&2.903304555&&&\\
\\
$2^3S$ & (ii)          &200&2.1749299&2.765&3.386&4.801\\
       & (iii)         &200&2.1749299&2.765&3.386&4.801\\
       & CE\cite{frolov86} &over 350&2.174930189&&&\\
\\
$2^1P$ & (ii)          &200&2.1235432&3.242&3.971&5.622\\
       & (iii)         &200&2.1235446&3.242&3.971&5.622\\
       & CE\cite{frolov86} &over 350&2.123545653&&&\\
\\
$2^3P$ & (ii)          &200&2.1328785&2.968&3.635&5.162\\
       & (iii)         &200&2.1328798&2.967&3.635&5.162\\
       & CE\cite{frolov86} &over 350&2.132880641&&&\\
\\
$3^1D$ & (ii)          &200&2.0553385&6.489&7.949&11.244\\
       & (iii)         &200&2.0553377&6.489&7.948&11.243\\
\\
       & (ii)$^a$      &200&2.0556201&6.488&7.948&11.242\\
       & (iii)$^a$     &200&2.0556195&6.488&7.947&11.242\\
       & Hylleraas \cite{sims}$^a$ &438&2.05562073279&&&\\
\\
$3^3D$ & (ii)          &200&2.0553538&6.486&7.945&11.239\\
       & (iii)         &200&2.0553531&6.486&7.945&11.238\\
\\
       & (ii)$^a$      &200&2.0556355&6.485&7.944&11.237\\
       & (iii)$^a$     &200&2.0556349&6.485&7.944&11.237\\
       & Hylleraas \cite{sims}$^a$ &393&2.05563630941&&&\\
\\
$4^1F$ & (ii)          &200&2.03097661&10.963&13.429&18.992\\
       & (iii)         &200&2.03097596&10.962&13.427&18.990\\
\\
       & (ii)$^a$      &200&2.03125504&10.962&13.427&18.989\\
       & (iii)$^a$     &200&2.03125314&10.960&13.425&18.987\\
       & Hylleraas \cite{sims}$^a$ &438&2.03125514434&&&\\
\\
$4^3F$ & (ii)          &200&2.03097664&10.963&13.429&18.992\\
       & (iii)         &200&2.03097598&10.962&13.427&18.990\\
\\
       & (ii)$^a$      &200&2.03125506&10.961&13.427&18.989\\
       & (iii)$^a$     &200&2.03125317&10.960&13.425&18.987\\
       & Hylleraas \cite{sims}$^a$ &438&2.03125516836&&&\\
\\
$5^1G$ & (ii)  &200&2.0197237802&16.589&20.321&28.738\\
       & (iii) &200&2.0197114443&16.553&20.276&28.675\\
\\
$5^3G$ & (ii)  &200&2.0197237803&16.589&20.321&28.738\\
       & (iii) &200&2.0197114445&16.553&20.276&28.675\\
\\
\hline
\end{tabular}
$^a$ The mass of the $\alpha$-particle is assumed to be infinite.

\end{table}


\begin{thebibliography}{99}
\bibitem{prca} K. Varga and Y.Suzuki, Phys. Rev. C{\bf 52}, 2885 (1995); 
Phys. Rev. A{\bf 53}, 1907 (1996). 
\bibitem{corrgauss} S. F. Boys, Proc. R. Soc. London Ser. A{\bf 258}, 
402 (1960);
 K. Singer, {\it ibid.} A{\bf 258}, 412 (1960).
\bibitem{temper} S. A. Alexander, H. J. Monkhorst, and K. Szalewicz,
J. Chem. Phys. {\bf 85}, 5821 (1986).
\bibitem{KA91}  P. M. Kozlowski and L. Adamowicz, J. Chem. Phys. {\bf 95}, 
6681 (1991).
\bibitem{KP93} D. B. Kinghorn and R. D. Poshusta, Phys. Rev. 
A{\bf 47}, 3671 (1993).
\bibitem{cencek} W. Cencek and J. Rychlewski, J. Chem. Phys. {\bf 98}, 
1252 (1993).
\bibitem{frolov96} A. M. Frolov and V. H. Smith, Jr., Phys. Rev. A{\bf 53}, 
3853 (1996) and references therein.
\bibitem{kamimura} M. Kamimura, Phys. Rev. A{\bf 38}, 621 (1988). 
\bibitem{kameyama} H. Kameyama, M. Kamimura, and Y. Fukushima, Phys. Rev. 
C{\bf 40}, 974 (1989).
\bibitem{vos} K. Varga, Y. Ohbayasi, and Y. Suzuki, to be published in 
Phys. Lett. B.
\bibitem{vinit} S. I. Vinitski${\breve {i}}$, V. S. Melezhik, 
L. I. Ponomarev, I. V. Puzynin, T. P. Puzynina, L. N. Somov, and 
N. F. Truskova, Sov. Phys. JETP {\bf 52}, 353 (1980).
\bibitem{bhatia} A. K. Bhatia and R. J. Drachman, Phys. Rev. A{\bf 30}, 2138 
(1984).
\bibitem{alexander} S. A. Alexander and H. J. Monkhorst, Phys. Rev. 
A{\bf 38}, 26 (1988).
\bibitem{frolov94} A. M. Frolov, V. H. Smith, Jr., and D. M. Bishop, 
Phys. Rev. A{\bf 49}, 1686 (1994); {\it ibid.} A{\bf 51}, 3636 (1995).  
\bibitem{frolov86} A. M. Frolov, Sov. Phys. JETP {\bf 65}, 1100 (1987). 
\bibitem{sims} J. S. Sims and W. C. Martin, Phys. Rev. A{\bf 37}, 2259 
(1988).
\bibitem{bukowski} See, for example, R. Bukowski, B. Jeziorski, S. Rybak, 
and K. Szalewicz, J. Chem. Phys. {\bf 102}, 888 (1995).
\bibitem{SVAO} K. Arai, Y. Suzuki, and K. Varga, Phys. Rev. C{\bf 51}, 
2488 (1995); K. Varga, Y. Suzuki, and I. Tanihata, {\it ibid}. C{\bf 52}, 
3013 (1995).
\bibitem{chong} D. P. Chong and D. M. Schrader, Mol. Phys. {\bf 16}, 
137 (1969).
\bibitem{SVM} V. I. Kukulin and V. M. Krasnopol'sky, J. Phys. G{\bf 3}, 795
(1977).
\bibitem{VSL} K. Varga, Y. Suzuki, and R. G. Lovas, Nucl. Phys. {\bf A571},
447 (1994).
\bibitem{ho} Y. K. Ho, J. Phys. B{\bf 16}, 1503 (1983).
\bibitem{bhatia83} A. K. Bhatia and R. J. Drachman, Phys. Rev. A{\bf 28}, 
2523 (1983).
\bibitem{yeremin} A. Yeremin, A. M. Frolov, and E. B. Kutukova, Few-Body 
Systems {\bf 4}, 111 (1988). 
\bibitem{krivec} R. Krivec, M. I. Haftel, and V. B. Mandelzweig, Phys. Rev. 
A{\bf 47}, 911 (1993).
\bibitem{ps-} A. M. Frolov, J. Phys. B{\bf 26}, 1031 (1993).
\bibitem{ho93} Y. K. Ho, Phys. Rev. A{\bf 48}, 4780 (1993).
\end{thebibliography}
\end{document}